\documentclass[aps,prb,twocolumn,showpacs,amsmath,amssymb,superscriptaddress,floatfix]{revtex4-2}

\usepackage[english]{babel}
\usepackage{blindtext}
\usepackage{latexsym}
\usepackage{amssymb}
\usepackage{physics}
\usepackage{amsmath}
\usepackage{bm}
\usepackage{amsfonts}
\usepackage{relsize}
\usepackage{xcolor}
\usepackage{verbatim}
\usepackage{bbold}
\usepackage{slashed}
\usepackage{appendix}
\usepackage{graphicx}
\usepackage{color}
\usepackage[normalem]{ulem}
\usepackage[colorlinks = true,
            linkcolor = blue,
            urlcolor  = blue,
            citecolor = blue,
            anchorcolor = blue]{hyperref}
\usepackage{graphicx} 
\usepackage{dcolumn} 
\newcommand{\approptoinn}[2]{\mathrel{\vcenter{
  \offinterlineskip\halign{\hfil$##$\cr
    #1\propto\cr\noalign{\kern2pt}#1\sim\cr\noalign{\kern-2pt}}}}}

\definecolor{ao(english)}{rgb}{0.0, 0.5, 0.0}
\definecolor{amaranth}{rgb}{0.9, 0.17, 0.31}
\definecolor{green(html/cssgreen)}{rgb}{0.0, 0.5, 0.0}

\newcommand\greensout{\bgroup\markoverwith{\textcolor{green(html/cssgreen)}{\rule[0.5ex]{2pt}{1.0pt}}}\ULon}

\begin{document}

\title{
Hall effect in topologically trivial isolated flat-band systems
}

\author{Raigo Nagashima}
\affiliation{Department of Physics, The University of Tokyo, Hongo, Tokyo, 113-8656, Japan}

\author{Masao Ogata}
\affiliation{Department of Physics, The University of Tokyo, Hongo, Tokyo, 113-8656, Japan}
\affiliation{Trans-scale Quantum Science Institute, The University of Tokyo, Hongo, Tokyo, 113-8656, Japan}

\author{Naoto Tsuji}
\affiliation{Department of Physics, The University of Tokyo, Hongo, Tokyo, 113-8656, Japan}
\affiliation{RIKEN Center for Emergent Matter Science (CEMS), Wako 351-0198, Japan}

\date{\today}

\begin{abstract}
We study the Hall effect in topologically trivial isolated flat-band systems (i.e., flat bands are separated from other bands and have zero Chern number) for a weak magnetic field.
In a naive semiclassical picture, the Hall conductivity vanishes when dispersive bands are unoccupied, since there are no mobile carriers.
To go beyond the semiclassical picture, we establish a fully quantum mechanical gauge-invariant formula for the Hall conductivity that can be applied to any lattice models.
We apply the formula to a general $N+M$-band model with $N$ dispersive bands and $M$-fold degenerate isolated flat bands, and find that when the dispersive bands are unoccupied, the total conductivity takes a universal form 
consisting of the energy difference between the dispersive and flat bands, and the non-Abelian quantum geometric tensor of the flat bands, which can be nonzero in systems with vanishing Berry curvature.
We numerically confirm the Hall effect for isolated flat-band lattice models on the honeycomb lattice ($N=M=1$) and two different Kagome lattices ($N=2$, $M=1$ and $N=1$, $M=2$).
\end{abstract}

\maketitle

\textit{Introduction}.---
Hall effect is a hallmark of transport phenomena \cite{Hall1879} that can bring us essential information on charge carriers in materials, including their density, mass, the sign of charges, and scattering properties. At weak magnetic fields, one can often adopt a semiclassical transport theory, in which the Hall conductivity is given by \cite{Kittel, Grosso_Parravicini}
\begin{align}
    \sigma_{xy}
    &=
    \frac{\sigma_0\omega_{\text{c}}\tau}{1+(\omega_{\text{c}}\tau)^2},
\end{align}
where $\sigma_0=ne^2\tau/m$, $\omega_{\text{c}}=eB/m$ is the cyclotron frequency, $n$, $e(>0)$, and $m$ are the density, charge, and mass of carriers, $B$ is the magnetic field strength, and $\tau$ is the relaxation time. If one naively takes the large mass limit ($m\to\infty$), which corresponds to a flat-band limit, the Hall conductivity vanishes. 
A similar result is expected in a high magnetic field in two dimensions, since a Chern number (which determines the quantized Hall conductance) must be zero in exactly flat-band lattice systems \cite{Chen2014}.
Our natural question here is whether the Hall conductivity always vanishes in flat-band systems and, if not, whether we can find any fingerprints of genuine quantum effects in the Hall conductivity that can survive even in a weak magnetic-field regime.

Flat-band systems have attracted much attention for a long time, in part because the kinetic energy of carriers is suppressed, and the interaction dominates overall physics.
In this context, flat-band systems have been discussed as a platform of, e.g., ferromagnetism \cite{Lieb1989, Mielke1991, Tasaki1992, Tasaki_Mielke1993, Tasaki1995}, fractional Chern insulators (for nearly flat bands) \cite{Tang2011, Sun2011, Neupert2011, Regnault2011, Bergholtz2013}, quantum anomalous Hall effects on flat bands \cite{Zhao2012, Zhou2022},
quantum geometric effects with quadratic band touching \cite{Rhim2020}, and flat-band superconductivity \cite{Misumi2017, Roy2019, Aoki2020, Nunes2020, Peri2021, Tian2023}.
Flat-band systems have been realized not only in crystalline solids \cite{Mizoguchi2019, Regnault2022, Neves2024} such as Kagome \cite{Kang2020}, pyrochlore \cite{Wakefield2023}, and Moir\'e \cite{Morell2010, Bistritzer2011, Chebrolu2019, Lisi2021} materials,
but also in magnon flat bands \cite{Schulenburg2002, Zhietomirsky2004, Tacchi2023}, photonic lattices \cite{Baboux2016, Leykam2018}, phononic crystals \cite{Ma2021}, and metal-organic frameworks \cite{Pan2023} (see also \cite{Liu2013, Yamada2016}).


\begin{figure}
    \centering
    \includegraphics[scale=0.6]{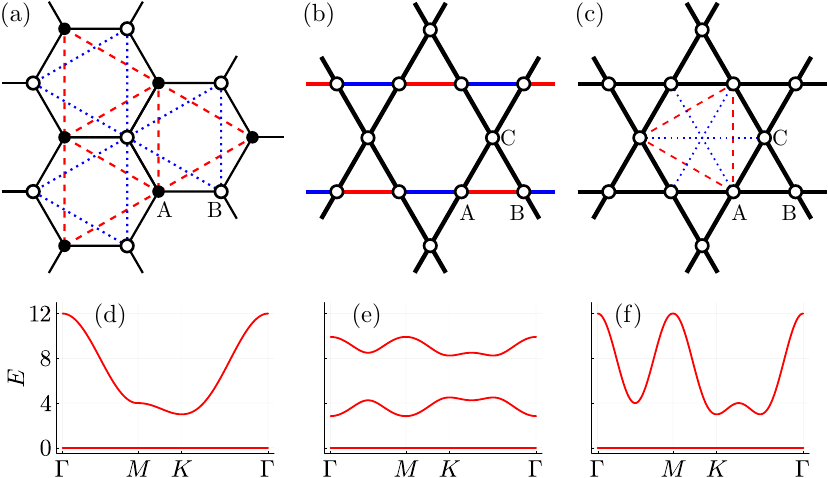}
    \caption{
    (a) Gapped honeycomb lattice model of nearest-neighbor and next-nearest-neighbor hoppings (black, red-dashed, and blue-dotted lines correspond to the hopping strength $\sqrt{3}t$, $t$, and $0$), having one isolated flat band at $E=0$~\cite{Misumi2017}.
    (b) Gapped Kagome lattice model with nearest-neighbor hoppings of different strengths (black, red, and blue lines correspond to the hopping strengths $t$, $\alpha t$, $t/\alpha$; $\alpha=2$), having one isolated flat band at $E=0$~\cite{Bilitewski2018}.
    (c) Kagome-$3$ lattice model with nearest, next-nearest, and next-next-nearest neighbor hoppings with the same amplitude $t$, having two-fold degenerate isolated flat bands at $E=0$~\cite{Bergman2008}.
    [(d),(e),(f)] Band structures of the models of (a), (b), and (c), respectively.
    Energy is measured in units of $t$.
    }
    \label{fig:1}
\end{figure}

In this letter, we study the Hall effect in isolated topologically trivial flat-band lattice systems (i.e., systems in which exactly flat bands are separated from other bands and have zero Chern number; see Fig.~\ref{fig:1}(d)-(f)) in a weak magnetic field.
We assume that the system has time-reversal symmetry when there is no magnetic field.
While the Hall conductivity can be evaluated quantum mechanically from the standard Kubo formula, it is not trivial to extract its weak magnetic-field limit on a lattice in a gauge invariant way \footnote{One way to understand this is that as one changes the strength of the magnetic field the magnetic Brillouin zone changes discontinuously, and in the weak field limit the size of the magnetic Brillouin zone diverges. For example, electrons on a square lattice in a uniform magnetic field show a fractal-like band structure known as the Hofstadter butterfly. Such a fine structure is smeared out if we introduce dissipation on a lattice system. }. 
Previously, the weak-field Hall conductivity has been obtained in a continuum space \cite{Fukuyama1969, Review_Bismuth, Koenye2021}.
Here we derive a fully quantum mechanical gauge-invariant formula for the Hall conductivity that is applicable to any lattice models.
On the other hand, the formula for the orbital magnetic susceptibility has been discussed since Landau and Peierls for a continuum space \cite{Peierls1933, Fukuyama1971, Ogata_Fukuyama2015, Ogata2017}, and recently an exact relation between the Hall conductivity and the orbital magnetic susceptibility has been found for a continuum space \cite{Ogata2024}. 
In the present paper, we establish the corresponding exact relation for lattice systems.
We note that the longitudinal conductivity of flat-band systems has been explored, see, e.g.,~\cite{Mitsherling2022, Mera2022, Huhtinen2023}.

We apply the formula to a general $N+M$-band model with $N$ dispersive and $M$-fold degenerate isolated flat bands.
The flat bands are topologically trivial (the Chern number is zero).
When the chemical potential is close to the isolated flat band(s), we find that the Hall conductivity takes a finite value proportional to the inverse of the relaxation rate.
This response takes a universal form consisting of the energy difference between the dispersive and isolated flat bands and the non-Abelian quantum geometric tensor~\cite{Ma2010, Ding2024} of the degenerate isolated flat bands.
The finite Hall conductivity is attributed to the effect of interband transitions, as expressed through the non-Abelian quantum geometric tensor.
These results are demonstrated for a gapped honeycomb lattice ($N=M=1$: Fig.~\ref{fig:1}(a))~\cite{Misumi2017}, gapped Kagome lattice ($N=2$, $M=1$: Fig.~\ref{fig:1}(b))~\cite{Bilitewski2018}, and Kagome-$3$ lattice ($N=1$, $M=2$: Fig.~\ref{fig:1}(c))\cite{Bergman2008} as concrete examples.
The last example has zero Berry curvature at any point in the Brillouin zone.
We also study the behavior of the Hall conductivity of the gapped honeycomb lattice when the isolated flat band is turned into a dispersive band.

\textit{Derivation of the formula}.---We consider a general electron system in the presence of a magnetic field $B$ along the $z$-axis.
We neglect spin-orbit couplings and Zeeman effects \footnote{
In principle, our approach can include the spin-orbit coupling.
However, our formulation for including the effect of an external magnetic field relies on the Peierls substitution, which introduces a phase arising from the magnetic field when an electron hops from one site to another.
When we introduce spin-orbit coupling, the term arising from an electron hopping intertwined with spin degrees of freedom is allowed.
Since this hopping term acquires an additional Peierls phase, the weak-field limit will be much more complicated.
Including the Zeeman effect should be straightforward. }.
Our argument can be applied to weakly disordered systems, where Anderson localization does not occur.
The Hall conductivity $\sigma_{xy}$ can be calculated from the Kubo formula as
\begin{equation}
    \sigma_{xy} = \lim_{\omega\to0}\frac{e^{2}}{\mathrm{i}\omega}\big( \Pi^{\text{R}}_{xy}(\omega) - \Pi^{\text{R}}_{xy}(0) \big),
    \label{KuboFormula}
\end{equation}
where $\Pi^{\text{R}}_{xy}(\omega)$ is an analytically continued ($\mathrm{i}\omega_{\lambda}\to\hbar\omega+\mathrm{i}0^{+}$) current-current correlation function $\Pi_{xy}(\mathrm{i}\omega_{\lambda})$ with $\omega_{\lambda}=2\pi\lambda k_{\text{B}}T$ being the Matsubara frequency: 
\begin{equation}
    \Pi_{xy}\qty(\mathrm{i}\omega_{\lambda}) = \frac{1}{\beta}\frac{1}{V}\sum_{n,\bm{k},\sigma}\text{Tr}\qty[ v_{\bm{k},Bx}G_{\bm{k},B+}v_{\bm{k},By}G_{\bm{k},B}].
    \label{CorrelationFunction}
\end{equation}
Here, $\beta=1/k_{\text{B}}T$ ($T$ is the temperature), $V$ is the volume of the system, $v_{\bm{k}, B\nu}$ ($\nu=x,y$) and $G_{\bm{k}, B(+)}$ are the velocity operator and the Matsubara Green's function in the presence of the magnetic field $B$, respectively.
We take the summation over $n$ for the fermionic Matsubara frequency $\varepsilon_{n}=\qty(2n+1)\pi k_{\text{B}}T$ ($n\in\mathbb{Z}$), and the trace is taken over all the Bloch bands.
We abbreviate $G_{B}\qty(\bm{k},\mathrm{i}\varepsilon_{n})$ and $G_{B}\qty(\bm{k},\mathrm{i}\varepsilon_{n}+\mathrm{i}\omega_{\lambda})$ as $G_{\bm{k}, B}$ and $G_{\bm{k}, B+}$, respectively.
Since we are interested in the weak field limit ($B\to 0$), we take the terms in the order of $B^{1}$ in the velocity operator and the Green's function \cite{Raoux2015}.

We begin with the real-space one-body Hamiltonian $H$ without a magnetic field: $H = \sum_{jk}t_{jk}\ket{j}\bra{k}$,
where $j$ and $k$ are site indices and $t_{jk}$ is assumed to be real.
The system preserves the time-reversal symmetry before applying the magnetic field $B$.
We include the effect of $B$ by the Peierls substitution: $H\to H_{B}=\sum_{jk}t_{jk}e^{\mathrm{i}\varphi_{jk}}\ket{j}\bra{k}$ with $\varphi_{jk} = -(e/\hbar)\int_{j}^{k}\bm{A}\qty(\bm{r})\cdot d\bm{l}$, where $\bm{A}$ is a vector potential of a static magnetic field.
To guarantee the gauge invariance of the formula, we expand the correlation function in terms of $B$ (not $\bm{A}$) with a trick used in Ref.~\cite{Chen2011, Savoie2012}. 
This is achieved by using the fact that the trace of the correlation function always yields a flux around sites $j,k,l$: $\Phi_{jkl}=\varphi_{jk}+\varphi_{kl}+\varphi_{lj}=-(e/\hbar)\int\int_{jkl}\bm{B}\cdot d\bm{S}$, which is manifestly gauge invariant.
By defining the Green's function $G_{B}$ and the velocity operator $v_{B\nu}$ ($\nu=x,y$) in the presence of the magnetic field as $G_{B}=(\mathrm{i}\varepsilon_{n}-H_{B})^{-1}$ and $v_{B\nu}=(1/\mathrm{i}\hbar)[\nu, H_{B}]$ with the position operator $\nu=\sum_{l}\nu_{l}\ket{l}\bra{l}$, we can calculate the correlation function while preserving the gauge invariance.
After performing the standard contour integral and analytic continuation and taking the limit of $\omega\to0$ in Eq.~(\ref{KuboFormula}), we obtain a gauge-invariant formula for the Hall conductivity: $\sigma_{xy}^{\text{Total}}=\sigma_{xy}^{\text{TR}}+\sigma_{xy}^{\text{TH}}$, $\sigma_{xy}^{\text{X}}=\int_{-\infty}^{\infty}d\varepsilon (-f'(\varepsilon)) \sigma_{xy}^{\text{X}}(\varepsilon)$ ($\text{X}=\text{TR},\text{TH}$), with
\begin{align}
    \frac{\sigma_{xy}^{\text{TR}}(\varepsilon)}{\sigma_{xy0}} &= -\frac{1}{N}\sum_{\bm{k}}\text{Im Tr}\big[ H_{\bm{k}}^{x}G^{\text{R}}_{\bm{k}}H_{\bm{k}}^{y}G^{\text{R}}_{\bm{k}}H_{\bm{k}}^{x}G^{\text{R}}_{\bm{k}}H_{\bm{k}}^{y}G^{\text{A}}_{\bm{k}} \notag \\
    &\quad + H_{\bm{k}}^{y}G^{\text{R}}_{\bm{k}}H_{\bm{k}}^{x}G^{\text{R}}_{\bm{k}}H_{\bm{k}}^{y}G^{\text{R}}_{\bm{k}}H_{\bm{k}}^{x}G^{\text{A}}_{\bm{k}} \notag \\
    &\quad +  \frac{1}{2}H_{\bm{k}}^{xy}\big( G^{\text{R}}_{\bm{k}}H_{\bm{k}}^{x}G^{\text{R}}_{\bm{k}}H_{\bm{k}}^{y}G^{\text{A}}_{\bm{k}} + G^{\text{R}}_{\bm{k}}H_{\bm{k}}^{y}G^{\text{R}}_{\bm{k}}H_{\bm{k}}^{x}G^{\text{A}}_{\bm{k}} \notag \\
    &\quad + G^{\text{R}}_{\bm{k}}H_{\bm{k}}^{x}G^{\text{A}}_{\bm{k}}H_{\bm{k}}^{y}G^{\text{R}}_{\bm{k}} + G^{\text{R}}_{\bm{k}}H_{\bm{k}}^{y}G^{\text{A}}_{\bm{k}}H_{\bm{k}}^{x}G^{\text{R}}_{\bm{k}} \notag \\
    &\quad + G^{\text{A}}_{\bm{k}}H_{\bm{k}}^{x}G^{\text{R}}_{\bm{k}}H_{\bm{k}}^{y}G^{\text{R}}_{\bm{k}} + G^{\text{A}}_{\bm{k}}H_{\bm{k}}^{y}G^{\text{R}}_{\bm{k}}H_{\bm{k}}^{x}G^{\text{R}}_{\bm{k}} \big) \big] , \label{eq:TR} \\
    \frac{\sigma_{xy}^{\text{TH}}(\varepsilon)}{\sigma_{xy0}} &= \frac{1}{N}\sum_{\bm{k}}\text{Im Tr}\big[ H_{\bm{k}}^{x}G^{\text{R}}_{\bm{k}}H_{\bm{k}}^{y}G^{\text{R}}_{\bm{k}}H_{\bm{k}}^{x}G^{\text{R}}_{\bm{k}}H_{\bm{k}}^{y}G^{\text{R}}_{\bm{k}} \notag \\
    +&\frac{1}{2} H_{\bm{k}}^{xy}G^{\text{R}}_{\bm{k}} \big(H_{\bm{k}}^{x}G^{\text{R}}_{\bm{k}}H_{\bm{k}}^{y}G^{\text{R}}_{\bm{k}} + H_{\bm{k}}^{y}G^{\text{R}}_{\bm{k}}H_{\bm{k}}^{x}G^{\text{R}}_{\bm{k}} \big) \big]. \label{eq:TH}
\end{align}

Here, $H_{\bm{k}}$ is the one-body Hamiltonian in $\bm{k}$-space without the magnetic field,
$\partial_{\nu}=\partial/\partial (k_{\nu}a)$, $a$ is the lattice constant, $l_{B}=\sqrt{\hbar/eB}$ is the magnetic length, $\sigma_{xy0}=2(e^{2}/h)\cdot(a/l_{B})^{2}\propto B^{1}$, $N$ is the number of unit cells, $H_{\bm{k}}^{\nu}=\partial_{\nu}H_{\bm{k}}$, $H_{\bm{k}}^{xy}=\partial_x \partial_y H_{\bm{k}}$,
$G^{\text{R}}_{\bm{k}}=(\varepsilon^{\text{R}} - H_{\bm{k}})^{-1}$ and $G^{\text{A}}_{\bm{k}}=(\varepsilon^{\text{A}} - H_{\bm{k}})^{-1}$ are the retarded and advanced Green's functions, $\varepsilon^{\text{R}}=\varepsilon + \mathrm{i}\Gamma$ and $\varepsilon^{\text{A}}=\varepsilon - \mathrm{i}\Gamma$ with $\Gamma$ being the relaxation rate, 
and $f(\varepsilon)=1/(e^{\beta(\varepsilon - \mu)}+1)$ is the Fermi distribution function.
$\sigma_{xy}^{\text{TR}}$ corresponds to the transport contribution of the Hall conductivity, which includes the results of the Boltzmann equation and reproduces the classical Drude results when considering the free electrons, having both the retarded and advanced Green's functions.
$\sigma_{xy}^{\text{TH}}$ is the thermodynamic contribution of the Hall conductivity arising from the current induced by the magnetic field \cite{Ogata2024}, which only has the retarded Green's functions.
Only electrons close to the Fermi surface can contribute to the transport and thermodynamic components.

Some technical comments are in order.
Although the scattering rate $\Gamma$ is phenomenologically introduced, it is dominantly determined by impurity scattering at low temperature, which is generally momentum-independent.
This does not alter the band dispersion because it lacks momentum dependence. \color{black}
We note that the $H^{xy}_{\bm{k}}$ terms have been absent in the previous research based on the continuous model~\cite{Fukuyama1969, Ogata2024, Nourafkan2018}.
These terms naively describe the effect of anisotropy of the energy dispersions.
We also note that the formulation in Ref.~\cite{Nourafkan2018} (which is similar to the one in Ref.~\cite{Fukuyama1969} and~\cite{Ogata2024}) does not properly account for the Peierls phase terms when calculating the non-interacting current vertex.

We can show that the thermodynamic contribution of the Hall effect $\sigma_{xy}^{\text{TH}}$ satisfies the following exact relation with the orbital magnetic susceptibility $\chi_{\rm orb}$ for lattice systems:
\begin{equation}
    \frac{\partial }{\partial \mu}\chi_{\text{orb}} = -\frac{ \sigma_{\text{orb}}^{\text{TH}}}{eB}.
    \label{exact_chi_sigma}
\end{equation}
Here the expression of  $\chi_{\text{orb}}$ for lattice systems has been shown in Ref.~\cite{Gomez-Santos2011, Raoux2015}. The relation (\ref{exact_chi_sigma}) is a straightforward extension of the one derived for a continuum space \cite{Ogata2024} (also known as the Str\v{e}da formula \cite{Streda1982, Widom1982, Streda1983}) to lattice systems.
Since our formula for $\sigma_{xy}^{\text{Total}}$ (Eqs.~(\ref{eq:TR}) and (\ref{eq:TH})) is gauge invariant, and the gauge invariance of the orbital magnetic susceptibility $\chi_{\text{orb}}$ guarantees the gauge invariance of $\sigma_{xy}^{\text{TH}}$ through Eq.~(\ref{exact_chi_sigma}), the gauge invariance of $\sigma_{xy}^{\text{TR}}$ is assured.
We note that the Hall response is zero when the external magnetic field vanishes, since both transport $\sigma_{xy}^{\text{TR}}$ and thermodynamic $\sigma_{xy}^{\text{TH}}$ contributions are proportional to the constant $\sigma_{xy0}=2(e^{2}/h)\cdot(a/l_{B})^{2}\propto B^{1}$.

\textit{General $N+M$-band model}.---We apply the formula to a $N+M$-band model with $N$ dispersive (with eigenfunctions $u_{n\bm{k}}$ and energy $E_{n\bm{k}}$ for $n\in D=\{1,\cdots,N\}$) and $M$-fold degenerate isolated flat bands ($u_{l\bm{k}}$ and $E_{l\bm{k}}=E_{\text{flat}}=\text{const.}$ for $l\in F=\{N+1,\cdots,N+M\}$), satisfying $|E_{n\in D,\bm{k}}-E_{\text{flat}}|\gg\Gamma>0$ for any $\bm{k}$-points in the Brillouin zone.
The degenerate isolated flat bands are topologically trivial in the sense that the Chern numbers are zero (where we use the non-Abelian Berry curvature for degenerate flat bands)~\footnote{We note that we consider the strong topology (independent of the spatial symmetry and characterized by the dimension of the whole system) to define the topologically trivial isolated flat bands. The trace of the non-Abelian Berry curvature vanishes in our setting, since the system preserves time-reversal symmetry (when the external magnetic field is absent) and lacks spin-orbit coupling; hence, we call the system topologically trivial. }, and their degeneracy does not arise from spin.
We take the band basis in which the Green's functions are diagonal: $G^{\text{R}/\text{A}}_{nn'}=(\varepsilon^{\text{R}/\text{A}}-E_{n\bm{k}})^{-1}\delta_{nn'}$.
After performing some calculations with the explicit expressions of $H^{\nu}_{\bm{k}}$ and $H^{xy}_{\bm{k}}$, we arrive at a simple description for the Hall conductivity around the degenerate isolated flat bands ($|\mu-E_{\text{flat}}|\ll\text{min}(|E_{n\in D,\bm{k}}-E_{\text{flat}}|)$) at $T=0$,
\begin{align}
    &\frac{\sigma^{\text{Total}}_{xy}(\mu)}{\sigma_{xy0}} = \frac{\Gamma^{3}}{[(\mu-E_{\text{flat}})^{2}+\Gamma^{2}]^{2}}\cdot\Lambda, \notag \\
    &\Lambda :=-4\cdot\frac{1}{N}\sum_{\bm{k}}\sum_{n\in D}\sum_{l,m\in F}(E_{n\bm{k}}-E_{\text{flat}}) \notag \\
    &\quad\times\text{Re}\left[ (Q^{lm}_{xy\bm{k},n}-Q^{lm}_{yx\bm{k},n})Q^{ml}_{xy\bm{k}}\right],
    \label{Around_flat_bands}
\end{align}
neglecting the terms that become zero at $\mu=E_{\text{flat}}$ when $\Gamma\to 0$. 
Here, $Q^{ab}_{xy\bm{k}}$ and $Q^{ab}_{xy\bm{k},n}$ is the full and the $n$-th component non-Abelian quantum geometric tensor~\cite{Ma2010}:
\begin{align}
    Q^{ab}_{xy\bm{k}} &= \sum_{j\in D}\braket{\partial_{x}u_{a\bm{k}}}{u_{j\bm{k}}}\braket{u_{j\bm{k}}}{\partial_{y}u_{b\bm{k}}} \notag \\
    &= \bra{\partial_{x}u_{a\bm{k}}}(1-\sum_{J\in F}\ket{u_{J\bm{k}}}\bra{u_{J\bm{k}}})\ket{\partial_{y}u_{b\bm{k}}}, \\
    Q^{ab}_{xy\bm{k},n} &= \braket{\partial_{x}u_{a\bm{k}}}{u_{n\bm{k}}}\braket{u_{n\bm{k}}}{\partial_{y}u_{b\bm{k}}},
\end{align}
satisfying $\sum_{n\in D}Q^{ab}_{xy\bm{k},n}=Q^{ab}_{xy\bm{k}}$.
The real part of the (non-Abelian) quantum geometric tensor describes how the state parametrized by the parameter (here $\bm{k}$) changes as the parameter infinitesimally changes, and its imaginary part gives the (non-Abelian) Berry curvature, corresponding to the (non-Abelian) gauge field.
For the finite temperature $T>0$ and exact clean limit ($\Gamma\to 0$), we replace $\Gamma^{3}/[(\mu-E_{\text{flat}})^{2}+\Gamma^{2}]^{2}$ with $(\pi/2)(\beta/4)\sech^{2}(\beta(\mu-E_{\text{flat}})/2)$.
See Supplemental Materials for the derivation.
We also show the individual expressions of the transport and thermodynamic contributions in Supplemental Materials.
$\Lambda$ does not diverge as the band gap $(E_{n\in D,\bm{k}}-E_{\text{flat}})$ grows, since the geometric part is suppressed as the band gap increases.
Importantly, $\Lambda$ can be nonzero even in the system preserving both inversion and time-reversal symmetry before applying an external magnetic field, where the Berry curvature is zero, as we see below.
We note that the non-Abelian quantum geometric tensor returns to the Abelian one when the isolated flat band does not have a degeneracy.
For the two-band case with one isolated flat band ($N=M=1$), the geometric part of the formula $\text{Re}[ (Q^{lm}_{xy\bm{k},n}-Q^{lm}_{yx\bm{k},n})Q^{ml}_{xy\bm{k}}]$ is reduced to the square of the Berry curvature of the isolated flat band, and its integral is non-vanishing even in the topologically trivial system.

The formula Eq.~(\ref{Around_flat_bands}) is universal in any system with isolated flat band(s), yielding a divergent response in the clean limit.
The form of Eq.~(\ref{Around_flat_bands}) shows a clear separation of the geometric contribution described by the non-Abelian quantum geometric tensor and the contribution from the band structure.
The Hall conductivity depends on the extrinsic parameter $\Gamma$ around the degenerate flat bands $\mu\approx E_{\text{flat}}$ as $\sigma_{xy}^{\text{Total}}\propto\Gamma^{-1}$, which differs from that of the dispersive bands depending on $\Gamma$ as $\sigma_{xy}^{\text{Total}}\propto\Gamma^{-2}$~\cite{Ogata2024}, indicating that the relaxation of the Hall response is suppressed more than in the dispersive bands.
The formula remains valid even when the isolated flat bands acquire finite dispersions and/or there is a small split between them, as long as they are smaller than $\Gamma$.

Since the longitudinal conductivity $\sigma_{\nu\nu}$ under no magnetic field in the clean limit neglecting the dispersive terms reads as $\sigma_{\nu\nu}=8(e^{2}/h)(1/N)\sum_{\bm{k},l\in F}Q^{ll}_{\nu\nu\bm{k}}$, which is positive  $Q^{ll}_{\nu\nu\bm{k}}=\sum_{j\in D, l\in F}|\braket{u_{j\bm{k}}}{\partial_{\nu}u_{l\bm{k}}}|^{2}>0$, the $B$ dependence of the resistivity $\rho_{xy}(B)=\sigma_{xy}^{\text{Total}}(B)/(\sigma_{xx}\sigma_{yy}+(\sigma_{xy}^{\text{Total}}(B))^{2})$ follows $\rho_{xy}(B)\propto B^{1}$ for small $B$.

The physical interpretation of the Hall conductivity is as follows: When the chemical potential is away from the isolated flat bands, there are no carriers, and the response is zero.
When the chemical potential is close to the isolated flat band (typically in the width of $2\Gamma$ or $2k_{\text{B}}T$ for finite $T$), electrons can be virtually excited from the isolated flat bands, where electrons become mobile and contribute to the Hall conductivity through geometric effects.
In the weak-field regime, the contribution is determined by the non-Abelian quantum geometric tensor, which is in contrast to the strong-field case of the dispersive bands, where the Berry curvature gives the quantized Hall conductivity, when the chemical potential is located between the Landau levels.

Although the interaction effect is not included here, our formula will be valid as long as the energy scale of the interaction is much smaller than that of the relaxation (or temperature).
For instance, when the isolated flat band is half-filled, its eigenfunction is spatially overlapping, and the system has onsite repulsive interactions $U>0$ at each site, one of the aligned spin configurations is chosen, and the ground state is ferromagnetic~\cite{Mielke1993, Mielke1999} with the energy gap proportional to $U$.
However, when the relaxation energy scale is much larger than $U$, the splitting of the flat bands is negligible and thus buried, implying that the system is effectively non-interacting; our formula remains valid.
This argument can be applied to the non-half-filling ferromagnetic case as well.

We note that our formalism can be applied to the system with ``nearly" degenerate isolated flat bands, where the flat bands are slightly gapped from each other, as long as the gap between the two flat bands is sufficiently smaller than the relaxation rate $\Gamma$.

\begin{figure}
    \centering
    \includegraphics[scale=0.5]{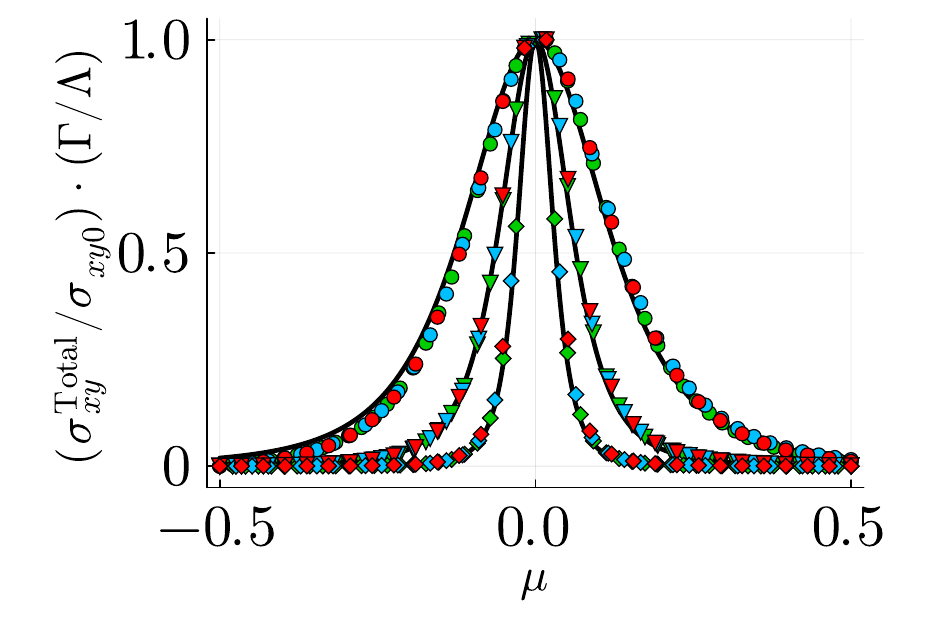}
    \caption{
    The normalized total Hall conductivity $(\sigma_{xy}^{\text{Total}}/\sigma_{xy0})\cdot(\Gamma/\Lambda)$ of the gapped honeycomb ($\Lambda\approx0.184$, red), gapped Kagome ($\Lambda\approx0.036$, blue), and Kagome-$3$ model ($\Lambda\approx6.357$, green).
    Eq.~(\ref{Around_flat_bands}) gives the solid lines for $\Gamma=0.2$, $0.1$, and $0.05$ from outside to inside. 
    The markers are calculated by Eqs.~(\ref{eq:TR}) and~(\ref{eq:TH}) for $\Gamma=0.2$ (circle), $0.1$ (triangle), and $0.05$ (diamond).
    A finite small value at $\mu=-0.5$ is subtracted in the markers.
    }
    \label{fig:2}
\end{figure}

\textit{Concrete examples.}---We consider the three concrete examples based on the lattice models in Fig.~\ref{fig:1}.
We fix the hopping amplitude $t=1$ as a unit of energy and consider the $T=0$ case for simplicity.
See Supplemental Material for the details of the models.
For the finite temperature case, we read $\Gamma$ as $(8/\pi)k_{\text{B}}T\sim k_{\text{B}}T$.
The first example is the gapped honeycomb lattice model with an isolated flat band~\cite{Misumi2017} (Fig.~\ref{fig:1}(a)), consisting of the nearest neighbor hopping $\sqrt{3}t$, the next-nearest neighbor hopping $t$ only from site $A$ to $A$, and the onsite potential $mt=3t$ at each site.
The second example is the gapped Kagome lattice model with an isolated flat band and two dispersive bands~\cite{Bilitewski2018} (Fig.~\ref{fig:1}(b)).
This model has three kinds of nearest-neighbor hopping amplitude: $t$ between sites A and C, and B and C; $\alpha t$ between sites A and B in one direction and $t/\alpha$ in the opposite, and the onsite potential $\alpha^{2}+(1/\alpha)^{2}$ at each site.
We choose $\alpha=2$ in the calculation.
The third example is the Kagome-$3$ model with two-fold degenerate isolated flat bands and one dispersive band~\cite{Bergman2008} (Fig.~\ref{fig:1}(c)).
This model includes the nearest, next-nearest, and next-next-nearest neighbor hoppings with the same amplitude $t$ and the onsite potential $2t$ at each site.
We note that the trace of the non-Abelian Berry curvature is zero, indicating the complete absence of the conventional topological effect.
All the models have (an) isolated flat band(s) at $E=0$.
The results of the numerical calculation of the Hall conductivity with $\mu\approx E_{\text{flat}}$ directly by Eq.~(\ref{Around_flat_bands}), and by Eqs.~(\ref{eq:TR}) and (\ref{eq:TH}) in Fig.~\ref{fig:2}, are represented by the solid lines and the markers, respectively.
The solid lines are calculated for $\Gamma=0.2$, $0.1$, and $0.05$ from outside to inside.
The red, green, and blue colors, and circle, triangle, and diamond markers correspond to the case of the gapped honeycomb ($\Lambda\approx0.184$), gapped Kagome ($\Lambda\approx0.036$), and Kagome-$3$ model ($\Lambda\approx6.357$), and $\Gamma=0.2$, $0.1$, and $0.05$, respectively.
Both methods yield results that agree with each other very well, so we can conclude that the total Hall conductivity at the isolated flat band(s) indicates the universal behavior described by the intertwined combination of the (non-)Abelian quantum geometric tensor and band-structure contributions in Eq.~(\ref{Around_flat_bands}).
The greatly enhanced Hall response in the Kagome-$3$ model ($\Lambda\approx6.357$) suggests a significant contribution from the degenerate isolated flat bands via the non-Abelian quantum geometric tensor, which cannot be captured by the conventional topological characterization by the (non-Abelian) Berry curvature.
See Supplemental Materials for the plots of the geometric part of the formula of those models.
The results presented here can also be applied to the finite temperature case, provided that the energy scale of the temperature, $k_{\text{B}}T$, is smaller than the energy gap between the dispersive and isolated flat bands. 

\begin{figure}
    \centering
    \includegraphics[scale=0.85]{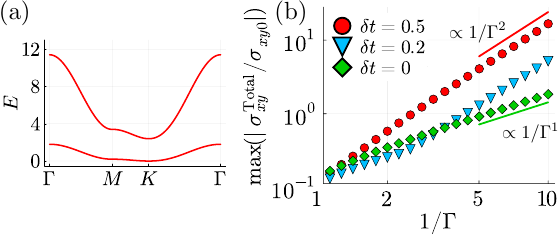}
    \caption{
    (a) The modified band structure of the gapped honeycomb lattice model with $\delta t=0.2$.
    (b) $\Gamma$-dependence of the maximum absolute value of the total Hall conductivity $|\sigma_{xy}^{\text{Total}}/\sigma_{xy0}|$ around the isolated flat band ($\mu\in[-t,t]$).
    The red circle, blue triangle, and green diamond curves correspond to the case of $\delta t=0.5$, $0.2$, and $0$, respectively.
    }
    \label{fig:3}
\end{figure}

\textit{Crossover}.---To understand the role of the flatness of the isolated flat band, we here introduce a finite dispersion to the isolated flat band of the gapped honeycomb lattice model without changing the form of the dispersive band.
The magnitude of the nearest-neighbor hopping, next-nearest neighbor hopping from sites A to A and B to B are $\sqrt{3}(t-\delta t)$, $t$, and $\delta t$, respectively.
The modified band structure with $\delta t=0.2$ is shown in Fig.~\ref{fig:3}(a).
The dispersive band shifts $-3\delta t$ and does not change the $\bm{k}$-dependence.

We calculate the Hall conductivity for this model by sweeping the parameter $\Gamma$ while fixing $\delta t$ within $\mu\in[-t,t]$, and obtain the $\Gamma$-dependence of the peak height of the total Hall conductivity ($\text{max}(|\sigma^{\text{Total}}_{xy}/\sigma_{xy0}|))$ as shown in Fig.~\ref{fig:3}(b).
The red circle, blue triangle, and green diamond curves correspond to the parameters $\delta t=0.5$, $0.2$, and $0$, respectively.
We can see a clear crossover from the flat band ($\propto\Gamma^{-1}$) to the dispersive one ($\propto\Gamma^{-2}$) in the blue triangle curve.
The crossover occurs around $\delta t/(\Gamma/2)\approx 1$, suggesting that the band width should be smaller than $\Gamma/2$ to realize the Hall response of the isolated flat band.
When $\delta t$ is smaller than $\Gamma/2$, the band can be seen as a flat band, and the Hall response is proportional to $\Gamma^{-1}$, exhibiting the intrinsic nature of the Hall effect.
On the other hand, when $(2\delta t)$ becomes larger than $\Gamma$ and the lower band becomes dispersive, the transport contribution overwhelms the thermodynamic contribution (see Supplemental Materials).
Moreover, the response becomes proportional to $\Gamma^{-2}$, a well-known property observed in the case of free electrons~\cite{Ogata2024}.
We note that $\sigma_{xy}^{\text{Total}}$ does not change its sign within $\mu\in[-t,t]$ when $\delta t$ is small (in our case, $\delta t=0$, $0.2$) and it is always positive.
When the dispersion effect becomes noticeable ($\delta t=0.5$), $\sigma_{xy}^{\text{Total}}$ changes its sign near $\mu=0$, which is consistent with the usual behavior of the Hall conductivity of the free electron~\cite{Ogata2024}.
This crossover behavior is somewhat similar to that of the longitudinal conductivity~\cite{Mitsherling2022}.
The modification of the formula (Eq.~(\ref{Around_flat_bands})) by introducing small dispersion to the isolated flat band for the two-band case ($N=M=1$) is presented in Supplemental Materials.

\textit{Relation to experiments}.---Let us discuss the observability of the phenomena in experiments.
In typical materials with the lattice constant $a\sim$ 1[\AA], the ratio $(a/l_B)^2$ is in the order of $10^{-5}$ for a typical magnetic-field strength $B\sim 1$ [T].
Then, the typical scale of the Hall conductivity in isolated flat-band systems is in the order of $10^{-5}\times (e^2/h)$, which is within the reach of experimental observations.
The observation will become much easier in clean samples because of the $\Gamma^{-1}$ dependence.
If one considers systems with a large lattice constant (e.g., Moire materials with $a\sim10^{-8}$ [m]), the value of the Hall conductivity can be enhanced.
A challenging part is finding appropriate materials with isolated flat bands with a sufficiently large energy gap.
Candidate materials include the twisted bilayer graphene \cite{Lisi2021}, metal-organic frameworks \cite{Pan2023, Liu2013, Yamada2016}, and several chain-type lattice materials \cite{Neves2024}.

\textit{Summary}.---We have investigated the Hall effect in isolated flat-band systems in the weak-field limit.
The derived gauge-invariant formula for the Hall conductivity [Eqs.~(\ref{eq:TR}) and~(\ref{eq:TH})], and the exact relation between the orbital magnetic susceptibility and the thermodynamic contribution of the Hall conductivity [Eq.~(\ref{exact_chi_sigma})] can be applied to any lattice models.
The Hall conductivity in isolated flat-band systems [Eq.~(\ref{Around_flat_bands})] shows the universal form consisting of the non-Abelian quantum geometric tensor (the geometric part) and energy difference between the dispersive and isolated flat bands (the band-structure part), the former of which arises from the interband effect.
This indicates the breakdown of the semiclassical theory based on the Boltzmann equation, and a quantum-mechanical treatment is essential to examine the interband effect in systems with isolated flat bands.
Future problems include addressing the effects of interactions, disorders, and strong magnetic fields that exceed the linear response regime, as well as exploring other physical quantities that display the effect of the non-Abelian quantum geometric tensor in isolated flat band systems.

We appreciate the useful comments from J\"{o}rg Schmalian, Alexander Mirlin, Kazuaki Takasan, Shohei Imai, Taiga Nakamoto, and Shuta Matsuura.
This work is supported by JST FOREST (Grant No.~JPMJFR2131) and JSPS KAKENHI (Grant No.~JP24H00191, JP25H01246, and JP25H01251).

\bibliography{main.bib}

\onecolumngrid

\section*{Details of the derivation of the gauge-invariant formula for the Hall conductivity}
In this section, we show the details of the derivation of the gauge-invariant formula for the Hall conductivity. To this end,
let us evaluate the correlation function $\Pi_{xy}(\mathrm{i}\omega_{\lambda})$ in the presence of the magnetic field $B$:
\begin{equation}
    \Pi_{xy}(\mathrm{i}\omega_{\lambda}) = \frac{1}{\beta}\frac{1}{V}\sum_{n,\bm{k},\sigma}\text{Tr}[v_{\bm{k},Bx}G_{\bm{k},B+}v_{\bm{k},By}G_{\bm{k},B}].
\end{equation}
See the main text about the definitions of the quantities.
When the magnetic field $B$ is applied to the system perpendicular to the $xy$-plane, its effect can be included via the Peierls substitution, and the Hamiltonian $H=\sum_{jk}t_{jk}\ket{j}\bra{k}$ becomes
\begin{equation}
    H_{B} = \sum_{jk}t_{jk}e^{\mathrm{i}\varphi_{jk}}\ket{j}\bra{k} \quad \text{with} \quad \varphi_{jk} = -\frac{e}{\hbar}\int_{j}^{k}\bm{A}(\bm{r})\cdot d\bm{l},
\end{equation}
where $e>0$, and $\bm{A}(\bm{r})$ being a vector potential of a static magnetic field.
Because of the trace of the correlation function, the path is always closed, and the gauge-invariant flux $\Phi_{jkl}$ always appears:
\begin{equation}
    \Phi_{jkl} = \varphi_{jk} + \varphi_{kl} + \varphi_{lj} = -\frac{e}{\hbar}\oint_{jkl}\bm{A}(\bm{r})\cdot d\bm{l} = -\frac{e}{\hbar}\int\int_{jkl}\bm{B}\cdot d\bm{S}.
\end{equation}
Its explicit expression in the Cartesian coordinate is given by \cite{Raoux2015}
\begin{equation}
    \Phi_{jkl} = +\frac{eB}{2\hbar}\left[(x_{j}-x_{l})(y_{l}-y_{k}) - (y_{j}-y_{l})(x_{l}-x_{k}) \right].
\end{equation}
We define the Matsubara Green's function with and without the magnetic field as $G_{\bm{k},B}=(\mathrm{i}\varepsilon_{n}-H_{\bm{k},B})^{-1}$ and $G_{\bm{k}}=(\mathrm{i}\varepsilon_{n}-H_{\bm{k}})^{-1}$.
To guarantee the gauge invariance, we define another Green's function $\tilde{G}_{\bm{k}}$ as $\tilde{G}_{\bm{k}, jk}=e^{\mathrm{i}\varphi_{jk}}G_{\bm{k}, jk}$ \cite{Chen2011, Savoie2012}.
The definition of $G_{\bm{k}}$: $(\mathrm{i}\varepsilon_{n}-H_{\bm{k}})^{-1}G_{\bm{k}}=1$, yields $(\mathrm{i}\varepsilon_{n}-H_{\bm{k},B})^{-1}\tilde{G}_{\bm{k}}=1-\mathcal{T}_{\bm{k}}$, with $\mathcal{T}_{\bm{k}, jk} = e^{\mathrm{i}\varphi_{jk}}\sum_{l}(e^{\mathrm{i}\Phi_{jkl}}-1)t_{jl}G_{\bm{k},lk}$.
The quantity $\mathcal{T}_{\bm{k}}$ corresponds to the correction term when the magnetic field is applied to the system.
Now, we can expand the Green's function with the magnetic field $G_{\bm{k}, B}$ about $B$ gauge-independently:
\begin{equation}
    G_{\bm{k}, B} = \left(\mathrm{i}\varepsilon_{n}- H_{\bm{k}, B} \right)^{-1} = \tilde{G}_{\bm{k}}\left(1-\mathcal{T}_{\bm{k}}\right)^{-1} = \tilde{G}_{\bm{k}}\sum_{n\geq0}\mathcal{T}_{\bm{k}}^{n}.
\end{equation}
Up to the first order of $B$, we can approximate the Green's function $G_{\bm{k}, B}$ as
\begin{equation}
    (G_{\bm{k}, B})_{jk} \approx G^{(0)}_{\bm{k},jk} + G^{(1)}_{\bm{k},jk} = e^{\mathrm{i}\varphi_{jk}}G_{\bm{k}, jk} + \mathrm{i}e^{\mathrm{i}\varphi_{jk}}\sum_{lm}\Phi_{lmk}G_{\bm{k}, jl}t_{lm}G_{\bm{k}, mk},
    \label{Green_first_order}
\end{equation}
where we used $\Phi\propto B$.
The velocity operators are given by
\begin{equation}
    \mathrm{i}\hbar v_{Bx} = \sum_{jk}e^{\mathrm{i}\varphi_{jk}}H^{x}_{jk}\ket{j}\bra{k}, \quad \mathrm{i}\hbar v_{By} = \sum_{jk}e^{\mathrm{i}\varphi_{jk}}H^{y}_{jk}\ket{j}\bra{k}.
\end{equation}
We can calculate the correlation function and take the first-order term of $B$ using Green's function (\ref{Green_first_order}).
The $B^{1}$ terms arising from the correlation function are shown below.
\begin{equation}
    \text{Tr}\left[v_{Bx}G_{B+}^{(0)}v_{By}G^{(0)}_{B}\right] + \text{Tr}\left[v_{Bx}G_{B+}^{(1)}v_{By}G^{(0)}_{B}\right] + \text{Tr}\left[v_{Bx}G_{B+}^{(0)}v_{By}G^{(1)}_{B}\right].
\end{equation}
Each term yields the following $B^{1}$ terms:
\begin{align}
    -\hbar^{2}\left[v_{Bx}G_{B+}^{(0)}v_{By}G^{(0)}_{B}\right] &\to \mathrm{i}\sum_{jklm}(\Phi_{jkl} + \Phi_{jlm})H^{x}_{jk}G_{+,kl}H^{y}_{lm}G_{mj}, \\
    -\hbar^{2}\left[v_{Bx}G_{B+}^{(1)}v_{By}G^{(0)}_{B}\right] &\to \mathrm{i}\sum_{jklmnp}\Phi_{mnl}H^{x}_{jk}G_{+,km}t_{mn}G_{+,nl}H^{y}_{lp}G_{pj}, \\
    -\hbar^{2}\left[v_{Bx}G_{B+}^{(0)}v_{By}G^{(1)}_{B}\right] &\to \mathrm{i}\sum_{jklmnp}\Phi_{npj}H^{x}_{jk}G_{+,kl}H^{y}_{lm}G_{mn}t_{np}G_{pj},
\end{align}
where the $\bm{k}$-dependence is abbreviated here for simplicity.
When performing the calculation, we use the relation for an operator $O$: $O^{\nu}=[\nu, H_{\bm{k}}]/(\mathrm{i}\hbar)$.
We can also use the practical relation about the derivative of Green's function when summarizing the calculation:
\begin{align}
    G^{\nu}_{\bm{k}} &= G_{\bm{k}}H_{\bm{k}}^{\nu}G_{\bm{k}}, \\
    G^{\nu\lambda}_{\bm{k}} &= G_{\bm{k}}H_{\bm{k}}^{\nu\lambda}G_{\bm{k}} + G_{\bm{k}}H_{\bm{k}}^{\nu}G_{\bm{k}}H_{\bm{k}}^{\lambda}G_{\bm{k}} + G_{\bm{k}}H_{\bm{k}}^{\lambda}G_{\bm{k}}H_{\bm{k}}^{\nu}G_{\bm{k}}.
\end{align}

To derive Eq.~(\ref{Around_flat_bands}), we need the explicit forms of $H_{\bm{k}}^{\nu}$ and $H_{\bm{k}}^{\nu\lambda}$.
We start from the eigenvalue equation
\begin{align}
    H_{\bm{k}}u_{l'\bm{k}} = E_{l'\bm{k}}u_{l'\bm{k}}.
\end{align}
By taking a derivative with respect to $\bm{k}$, we obtain
\begin{equation}
(\nabla_{\bm{k}}H_{\bm{k}})u_{l'\bm{k}} + H_{\bm{k}}(\nabla_{\bm{k}}u_{l'\bm{k}}) = (\nabla_{\bm{k}}E_{l'\bm{k}})u_{l'\bm{k}} + E_{l'\bm{k}}(\nabla_{\bm{k}}u_{l'\bm{k}}).
\end{equation}
This leads to
\begin{equation}
\bra{ u_{l'\bm{k}} }\nabla_{\bm{k}}H_{\bm{k}} \ket{ u_{l'\bm{k}} } = (\nabla_{\bm{k}}E_{l\bm{k}})\delta_{ll'} - (E_{l\bm{k}} - E_{l'\bm{k}})\braket{ u_{l\bm{k}} }{\nabla_{\bm{k}}u_{l'\bm{k}} }.
\end{equation}
We arrive at
\begin{align}
    H^{\nu}_{\bm{k}} &= \mqty[ 
    H^{\nu}_{11\bm{k}} & H^{\nu}_{12\bm{k}} & \cdots & H^{\nu}_{1N+M\bm{k}} \\ 
    H^{\nu}_{21\bm{k}} & H^{\nu}_{22\bm{k}} & \cdots & H^{\nu}_{2N+M\bm{k}} \\
    \vdots & \vdots & \ddots & \vdots \\
    H^{\nu}_{N+M1\bm{k}} & H^{\nu}_{N+M2\bm{k}} & \cdots & H^{\nu}_{N+MN+M\bm{k}}], \\
    H^{\nu}_{ll'\bm{k}} &= (\partial_{\nu}E_{l\bm{k}})\delta_{ll'} - (E_{l\bm{k}} - E_{l'\bm{k}})\braket{u_{l\bm{k}}}{\partial_{\nu}u_{l'\bm{k}}},
\end{align}
with $\nu=x,y$ and $H^{\nu}_{ll'\bm{k}}=0$ for $l,l'\in F$.
We can derive the matrix elements of the second derivative of the Hamiltonian as follows.
\begin{align}
H^{xy}_{\bm{k}} &= \mqty[ 
    H^{xy}_{11\bm{k}} & H^{xy}_{12\bm{k}} & \cdots & H^{xy}_{1N+M\bm{k}} \\ 
    H^{xy}_{21\bm{k}} & H^{xy}_{22\bm{k}} & \cdots & H^{xy}_{2N+M\bm{k}} \\
    \vdots & \vdots & \ddots & \vdots \\
    H^{xy}_{N+M1\bm{k}} & H^{xy}_{N+M2\bm{k}} & \cdots & H^{xy}_{N+MN+M\bm{k}}], \\
    H^{xy}_{ll'\bm{k}} &= (\partial_{x}\partial_{y}E_{l\bm{k}})\delta_{ll'} - (E_{l\bm{k}} - E_{l'\bm{k}})\braket{u_{l\bm{k}}}{\partial_{x}\partial_{y}u_{l'\bm{k}}} \notag \\
    &\quad -\sum_{j}(E_{l\bm{k}} - E_{j\bm{k}})(\braket{\partial_{x}u_{l\bm{k}}}{u_{j\bm{k}}} \braket{u_{j\bm{k}}}{\partial_{y}u_{l'\bm{k}}} + \braket{\partial_{y}u_{l\bm{k}}}{u_{j\bm{k}}} \braket{u_{j\bm{k}}}{\partial_{x}u_{l'\bm{k}}}),
\end{align}
and $H^{xy}_{ll'\bm{k}}=-\sum_{j\in D}(E_{l\bm{k}} - E_{j\bm{k}})(\braket{\partial_{x}u_{l\bm{k}}}{u_{j\bm{k}}} \braket{u_{j\bm{k}}}{\partial_{y}u_{l'\bm{k}}} + \braket{\partial_{y}u_{l\bm{k}}}{u_{j\bm{k}}} \braket{u_{j\bm{k}}}{\partial_{x}u_{l'\bm{k}}})$ holds for $l,l'\in F$.\\

\noindent
We can formally rewrite the formulae Eqs.~(\ref{eq:TR}) and~(\ref{eq:TH}) by using the following equations.
\begin{align}
    \text{Tr}[H^{x}_{\bm{k}}G^{\text{R}}_{\bm{k}}H^{y}_{\bm{k}}G^{\text{R}}_{\bm{k}}H^{x}_{\bm{k}}G^{\text{R}}_{\bm{k}}H^{y}_{\bm{k}}G^{\text{A}}_{\bm{k}}] &= \sum_{jlmn}G^{\text{R}}_{l\bm{k}}G^{\text{R}}_{m\bm{k}}G^{\text{R}}_{n\bm{k}}G^{\text{A}}_{j\bm{k}}H^{x}_{jl\bm{k}}H^{y}_{lm\bm{k}}H^{x}_{mn\bm{k}}H^{y}_{nj\bm{k}}, \\
    \text{Tr}[H^{xy}_{\bm{k}}G^{\text{R}}_{\bm{k}}H^{y}_{\bm{k}}G^{\text{A}}_{\bm{k}}H^{x}_{\bm{k}}G^{\text{R}}_{\bm{k}}] &= \sum_{lmn}G^{\text{R}}_{m\bm{k}}G^{\text{A}}_{n\bm{k}}G^{\text{R}}_{l\bm{k}}H^{xy}_{lm\bm{k}}H^{y}_{mn\bm{k}}H^{x}_{nl\bm{k}}, \\
    \text{Tr}[H^{xy}_{\bm{k}}G^{\text{A}}_{\bm{k}}H^{x}_{\bm{k}}G^{\text{R}}_{\bm{k}}H^{y}_{\bm{k}}G^{\text{R}}_{\bm{k}}] &= \sum_{lmn}G^{\text{A}}_{m\bm{k}}G^{\text{R}}_{n\bm{k}}G^{\text{R}}_{l\bm{k}}H^{xy}_{lm\bm{k}}H^{x}_{mn\bm{k}}H^{y}_{nl\bm{k}}, \\
    \text{Tr}[H^{xy}_{\bm{k}}G^{\text{R}}_{\bm{k}}H^{x}_{\bm{k}}G^{\text{R}}_{\bm{k}}H^{y}_{\bm{k}}G^{\text{A}}_{\bm{k}}] &= \sum_{lmn}G^{\text{R}}_{m\bm{k}}G^{\text{R}}_{n\bm{k}}G^{\text{A}}_{l\bm{k}}H^{xy}_{lm\bm{k}}H^{x}_{mn\bm{k}}H^{y}_{nl\bm{k}}, \\
    \text{Tr}[H^{x}_{\bm{k}}G^{\text{R}}_{\bm{k}}H^{y}_{\bm{k}}G^{\text{R}}_{\bm{k}}H^{x}_{\bm{k}}G^{\text{R}}_{\bm{k}}H^{y}_{\bm{k}}G^{\text{R}}_{\bm{k}}] &= \sum_{jlmn}G^{\text{R}}_{l\bm{k}}G^{\text{R}}_{m\bm{k}}G^{\text{R}}_{n\bm{k}}G^{\text{R}}_{j\bm{k}}H^{x}_{jl\bm{k}}H^{y}_{lm\bm{k}}H^{x}_{mn\bm{k}}H^{y}_{nj\bm{k}}, \\
    \text{Tr}[H^{xy}_{\bm{k}}G^{\text{R}}_{\bm{k}}H^{x}_{\bm{k}}G^{\text{R}}_{\bm{k}}H^{y}_{\bm{k}}G^{\text{R}}_{\bm{k}}] &= \sum_{lmn}G^{\text{R}}_{m\bm{k}}G^{\text{R}}_{n\bm{k}}G^{\text{R}}_{l\bm{k}}H^{xy}_{lm\bm{k}}H^{x}_{mn\bm{k}}H^{y}_{nl\bm{k}},
\end{align}
where $G^{\text{R/A}}_{l\bm{k}} = (\varepsilon^{\text{R}/\text{A}} - E_{l\bm{k}})^{-1}$.
Since we consider the region close to the degenerate isolated flat bands, we approximate the Green's functions as
\begin{align}
    G^{\text{R}}_{l\bm{k}} &= -\frac{ (E_{l\bm{k}} - E_{\text{flat}}) + \mathrm{i}\Gamma }{(E_{l\bm{k}} - E_{\text{flat}})^{2}} \quad (l\in D), \\
    G^{\text{A}}_{l\bm{k}} &= -\frac{ (E_{l\bm{k}} - E_{\text{flat}}) - \mathrm{i}\Gamma }{(E_{l\bm{k}} - E_{\text{flat}})^{2}} \quad (l\in D), \\
    G^{\text{R}}_{l\bm{k}} &= -\frac{\mathrm{i}\Gamma}{\varepsilon^{2} + \Gamma^{2}} \quad (l\in F), \\
    G^{\text{A}}_{l\bm{k}} &= +\frac{\mathrm{i}\Gamma}{\varepsilon^{2} + \Gamma^{2}} \quad (l\in F).
\end{align}
We note that the eigenvalues of the system satisfy $E_{l,\bm{k}}=E_{l,-\bm{k}}$, as we do not consider spin-orbit coupling, and that the imaginary part of the products of $H^{\nu}_{\bm{k}}$ and $H^{xy}_{\bm{k}}$ vanish by integration over the Brillouin zone since the system is assumed to preserve the time-reversal symmetry except for the case with an external magnetic field (the useful example is the Berry curvature in the single-band case, which is an odd function of $\bm{k}$ in the system preserving the time-reversal symmetry).
After some calculations and taking the dominant terms, which are $O(\Gamma^{-1})$ at $\mu=E_{\text{flat}}$, we obtain Eq.~(\ref{Around_flat_bands}).\\

The individual expressions of ``transport" and ``thermodynamic" contributions are given as follows.
\begin{align}
\frac{\sigma_{xy}^{\text{TR}}(\varepsilon) }{\sigma_{xy0} } &= -\frac{1}{N}\sum_{\bm{k}}\Bigg( 4\sum_{l\in F}\sum_{jmn\in D}\frac{1}{\Delta_{j}\Delta_{m}\Delta_{n}}\frac{\Gamma}{\varepsilon^{2}+\Gamma^{2}}\text{Re}\left[ H_{jl}^{x}H_{lm}^{y}H_{mn}^{x}H_{nj}^{y} \right] \notag \\
&\quad + 4\sum_{jm\in F}\sum_{ln\in D}\frac{\Delta_{n}}{\Delta_{l}^{2}\Delta_{n}^{2} }\frac{\Gamma^{3}}{(\varepsilon^{2}+\Gamma^{2})^{2}}\text{Re}\left[ H_{jl}^{x}H_{lm}^{y}H_{mn}^{x}H_{nj}^{y} \right] \notag \\
&\quad - \sum_{n\in F}\sum_{lm\in D}\frac{1}{\Delta_{l}\Delta_{m}}\frac{\Gamma}{\varepsilon^{2}+\Gamma^{2}}\text{Re}\left[ H_{ln}^{xy}H_{nm}^{x}H_{ml}^{y} \right] \notag \\
&\quad - \sum_{l\in F}\sum_{mn\in D}\frac{1}{\Delta_{m}\Delta_{n}}\frac{\Gamma}{\varepsilon^{2}+\Gamma^{2}}\text{Re}\left[ H_{ln}^{xy}H_{nm}^{x}H_{ml}^{y} \right] \notag \\
&\quad - \sum_{m\in F}\sum_{ln\in D}\frac{1}{\Delta_{l}\Delta_{n}}\frac{\Gamma}{\varepsilon^{2}+\Gamma^{2}}\text{Re}\left[ H_{nl}^{xy}H_{lm}^{x}H_{mn}^{y} \right] \notag \\
&\quad -3\sum_{ln\in F}\sum_{m\in D}\frac{1}{\Delta_{m}^{2}}\frac{\Gamma^{3}}{(\varepsilon^{2}+\Gamma^{2})^{2}}\text{Re}\left[ H_{ln}^{xy}H_{nm}^{x}H_{ml}^{y} \right] \Bigg),
\end{align}
\begin{align}
\frac{\sigma_{xy}^{\text{TH}}(\varepsilon) }{\sigma_{xy0} } &= +\frac{1}{N}\sum_{\bm{k}}\Bigg( 4\sum_{l\in F}\sum_{jmn\in D}\frac{1}{\Delta_{j}\Delta_{m}\Delta_{n}}\frac{\Gamma}{\varepsilon^{2}+\Gamma^{2}}\text{Re}\left[ H_{jl}^{x}H_{lm}^{y}H_{mn}^{x}H_{nj}^{y} \right] \notag \\
&\quad -4\sum_{jm\in F}\sum_{ln\in D}\frac{\Delta_{n}}{\Delta_{l}^{2}\Delta_{n}^{2}}\frac{\Gamma^{3}}{(\varepsilon^{2}+\Gamma^{2})^{2}}\text{Re}\left[ H_{jl}^{x}H_{lm}^{y}H_{mn}^{x}H_{nj}^{y} \right] \notag \\
&\quad - \sum_{n\in F}\sum_{lm\in D}\frac{1}{\Delta_{l}\Delta_{m}}\frac{\Gamma}{\varepsilon^{2}+\Gamma^{2}}\text{Re}\left[ H_{nl}^{xy}H_{lm}^{x}H_{mn}^{y} \right] \notag \\
&\quad - \sum_{l\in F}\sum_{mn\in D}\frac{1}{\Delta_{m}\Delta_{n}}\frac{\Gamma}{\varepsilon^{2}+\Gamma^{2}}\text{Re}\left[ H_{nl}^{xy}H_{lm}^{x}H_{mn}^{y} \right] \notag \\
&\quad - \sum_{m\in F}\sum_{ln\in D}\frac{1}{\Delta_{l}\Delta_{n}}\frac{\Gamma}{\varepsilon^{2}+\Gamma^{2}}\text{Re}\left[ H_{nl}^{xy}H_{lm}^{x}H_{mn}^{y} \right] \notag \\
&\quad + \sum_{ln\in F}\sum_{m\in D}\frac{1}{\Delta_{m}^{2}}\frac{\Gamma^{3}}{(\varepsilon^{2}+\Gamma^{2})^{2}}\text{Re}\left[ H_{nl}^{xy}H_{lm}^{x}H_{mn}^{y} \right] \Bigg),
\end{align}
where we only take the dominant terms, and the $\bm{k}$-dependence is abbreviated here.
By summing both terms, several terms cancel out and the expression of the total contribution $\sigma_{xy}^{\text{Total}}$ becomes simpler:
\begin{align}
    \frac{\sigma_{xy}^{\text{Total}}(\varepsilon) }{\sigma_{xy0}} &= -8\frac{1}{N}\sum_{\bm{k}}\sum_{jm\in F}\sum_{ln\in D}\frac{\Gamma^{3}}{(\varepsilon^{2} + \Gamma^{2})^{2}}\frac{\Delta_{n}}{\Delta_{l}^{2}\Delta_{n}^{2}}\text{Re}\left[ H^{x}_{jl}H^{y}_{lm}H^{x}_{mn}H^{y}_{nj} \right] \notag \\
& + 4\frac{1}{N}\sum_{\bm{k}}\sum_{ln\in F}\sum_{m\in D}\frac{\Gamma^{3}}{(\varepsilon^{2}+\Gamma^{2})^{2}}\frac{1}{\Delta_{m}^{2}}\text{Re}\left[ H^{xy}_{nl}H^{x}_{lm}H^{y}_{mn} \right].
\end{align}
Since we have
\begin{align}
H^{x}_{jl}H^{y}_{lm}H^{x}_{mn}H^{y}_{nj} &= \Delta_{l}^{2}\Delta_{n}^{2}\braket{\partial_{x}u_{j}}{u_{l}}\braket{u_{l}}{\partial_{y}u_{m}}\braket{\partial_{x}u_{m}}{u_{n}}\braket{u_{n}}{\partial_{y}u_{j}} \quad (l,n\in D,\ j,m\in F) \\
H^{xy}_{nl}H^{x}_{lm}H^{y}_{mn} &= \Delta_{m}^{2}\braket{\partial_{x}u_{l}}{u_{m}}\braket{u_{m}}{\partial_{y}u_{n}} \notag \\
&\quad\quad \times \sum_{j\in D}\Delta_{j}\left[ \braket{\partial_{x}u_{n}}{u_{j}}\braket{u_{jn}}{\partial_{y}u_{l}} + \braket{\partial_{y}u_{j}}{u_{j}}\braket{u_{j}}{\partial_{x}u_{l}} \right] \quad (m\in D,\ l,n\in F),
\end{align}
the sum can be rewritten as below.
\begin{equation}
\frac{\sigma^{\text{Total}}_{xy} (\varepsilon)}{\sigma_{xy0}} = -4\frac{1}{N}\sum_{\bm{k}}\sum_{jm\in F}\sum_{ln\in D}\Delta_{n}\frac{\Gamma^{3}}{(\varepsilon^{2}+\Gamma^{2})^{2}}\text{Re}\left[ \braket{\partial_{x}u_{j}}{u_{l}}\braket{u_{l}}{\partial_{y}u_{m}} \left( \braket{\partial_{x}u_{m}}{u_{n}}\braket{u_{n}}{\partial_{y}u_{j}} - \braket{\partial_{x}u_{m}}{u_{n}}\braket{u_{n}}{\partial_{y}u_{j}} \right) \right].
\end{equation}
We introduce the following quantities:
\begin{align}
Q^{lm}_{xy,n}(\bm{k}) &= \braket{ \partial_{x}u_{l\bm{k}} }{ u_{n\bm{k}} } \braket{ u_{n\bm{k}} }{ \partial_{y}u_{m\bm{k}} }, \\
Q^{ml}_{xy}(\bm{k}) &= \sum_{j\in D} \braket{ \partial_{x}u_{m\bm{k}} }{ u_{j\bm{k}} } \braket{ u_{j\bm{k}} }{ \partial_{y}u_{l\bm{k}} } \\
&= \braket{\partial_{x}u_{m\bm{k}}}{\partial_{y}u_{l\bm{k}}} - \sum_{J\in F}\braket{\partial_{x}u_{m\bm{k}}}{u_{J\bm{k}}}\braket{u_{J\bm{k}}}{\partial_{y}u_{l\bm{k}}}, \\
\sum_{n\in D}Q^{lm}_{xy,n}(\bm{k}) &= Q^{ml}_{xy}(\bm{k}).
\end{align}
$Q^{lm}_{xy,n}$ is the $n$-th component of the non-Abelian quantum geometric tensor of the degenerate isolated flat bands ($n\in D$), and $Q^{ml}_{xy}$ is the full non-Abelian quantum geometric tensor of the degenerate isolated flat bands.
Then, we obtain the following formula.
\begin{equation}
\sigma_{xy}(\varepsilon) = \sigma_{xy0}\cdot (-4)\frac{1}{N}\sum_{\bm{k}}\sum_{n\in D}\sum_{l,m\in F}\frac{\Gamma^{3}}{(\varepsilon^{2}+\Gamma^{2})^{2}}(E_{n}(\bm{k}) - E_{\text{F}})\text{Re}\left[ \left( Q^{lm}_{xy,n}(\bm{k}) - Q^{lm}_{yx,n}(\bm{k}) \right) Q^{ml}_{xy}(\bm{k}) \right].
\end{equation}
For the finite-temperature case, we consider the convolution of the above function with the Fermi distribution.
Then, in the clean limit $\Gamma\to 0$, we alter the $\Gamma$-dependent part as
\begin{equation}
    \frac{\Gamma^{3}}{(\varepsilon^{2}+\Gamma^{2})^{2}}\to \frac{\pi}{2}\frac{1}{4k_{\text{B}}T}\sech^{2}\left(\frac{\mu}{2k_{\text{B}}T}\right).
\end{equation}
The relaxation rate $\Gamma$ has a relationship with the temperature $T$ in our case as
\begin{equation}
    \Gamma\to\frac{8}{\pi}k_{\text{B}}T\sim2k_{\text{B}}T.
\end{equation}

\section*{Model descriptions}
We show the details of the lattice models in the main part of the paper.\\

\noindent
\textit{Gapped honeycomb model}.---The first example is the gapped honeycomb model in Fig.~1(a), which has one isolated flat band and one dispersive band.
The Hamiltonian $H_{\text{gh}\bm{k}}$ is given by the following.
\begin{equation}
    H_{\text{gh}\bm{k}} = \mqty[
    t(F_{\text{k}}+m) & \sqrt{3}t\tilde{F}_{\bm{k}} \\
    \sqrt{3}t\tilde{F}^{*}_{\bm{k}} & tm
    ],
\end{equation}
where $F_{\bm{k}} = 4\cos(\sqrt{3}k_{x}a/2)\cos(k_{y}a/2) + 2\cos(k_{y}a)$ and $\tilde{F}_{\bm{k}}=e^{\mathrm{i}k_{x}a/\sqrt{3}}+2e^{-\mathrm{i}\sqrt{3}k_{x}a/6}\cos(k_{y}a/2)$, satisfying $|\tilde{F}_{\bm{k}}|^{2}=F_{\bm{k}}+3$, and $m$ is the on-site potential.
The model has nearest-neighbor and next-nearest-neighbor hoppings \cite{Misumi2017}.
The magnitude of the nearest-neighbor hopping, next nearest-neighbor hoppings from sites $A$ to $A$ and $B$ to $B$ are $\sqrt{3}t$, $t$, and $0$, respectively. 
The eigenvalues of this model are $E_{\text{gh},1{\bm k}}=t\qty(F_{\bm{k}}+3+m)$ and $E_{\text{gh},2{\bm k}}=t(m-3)$.
Since $m$ shifts the origin of the energy, there is essentially no free parameter in the model, and we cannot freely change the band width $(E_{\text{gh},1\bm{k}}-E_{\text{gh},2\bm{k}})$.\\

\noindent
When we introduce a finite dispersion to the isolated flat band of the honeycomb lattice model, without changing the form of the dispersive band, we modify the Hamiltonian $H_{\text{gh}{\bm k}}\to\tilde{H}_{\text{gh}{\bm k}}$ as
\begin{equation}
    \tilde{H}_{\text{gh}{\bm k}}=\mqty[ t(F_{\bm{k}}+m) & \sqrt{3}(t-\delta t)\tilde{F}_{\bm{k}} \\ \sqrt{3}\qty(t-\delta t)\tilde{F}_{\bm{k}}^{*} & \qty(\delta t)F_{\bm{k}} + tm ],
\end{equation}
where $\delta t$ controls the dispersion of the isolated lower band.
The magnitude of the nearest-neighbor hopping, next nearest-neighbor hoppings from sites $A$ to $A$ and $B$ to $B$ are $\sqrt{3}(t-\delta t)$, $t$, and $\delta t$, respectively.
The eigenvalues of this model are given by $\tilde{E}_{\text{gh},1{\bm k}}=t(F_{\bm{k}}+3+m) - 3\delta t$ and $\tilde{E}_{\text{gh},2{\bm k}}=t(m-3) + \delta t(F_{\bm{k}}+3)$, where the dispersive band $\tilde{E}_{\text{gh},1{\bm k}}$ is just shifted by $-3\delta t$ from the original one $E_{\text{gh},1{\bm k}}$. \\

\noindent
\textit{Gapped Kagome model}.---The second example is the gapped Kagome model in Fig.~1(b)~\cite{Bilitewski2018}, which has one isolated flat band and two dispersive bands.
The Hamiltonian $H_{\text{gk}\bm{k}}$ is given by the following.
\begin{equation}
    H_{\text{gk}\bm{k}} = t\cdot\mqty[
    \alpha^{2}+(1/\alpha)^{2} & \alpha^{2}e^{-\mathrm{i}\bm{k}\cdot\bm{\delta}_{\text{AB}}} + (1/\alpha)^{2}e^{\mathrm{i}\bm{k}\cdot\bm{\delta}_{\text{AB}}} & 2\cos(\bm{k}\cdot\bm{\delta}_{\text{CA}}) \\
    \alpha^{2}e^{\mathrm{i}\bm{k}\cdot\bm{\delta}_{\text{AB}}} + (1/\alpha)^{2}e^{-\mathrm{i}\bm{k}\cdot\bm{\delta}_{\text{AB}}} & \alpha^{2} + (1/\alpha)^{2} & 2\cos(\bm{k}\cdot\bm{\delta}_{\text{BC}}) \\
    2\cos(\bm{k}\cdot\bm{\delta}_{\text{CA}}) & 2\cos(\bm{k}\cdot\bm{\delta}_{\text{BC}}) & \alpha^{2} + (1/\alpha)^{2}
    ],
\end{equation}
where $\bm{\delta}_{\text{AB}}=[a,0]^{\text{T}}$, $\bm{\delta}_{\text{BC}} = [-a/2,\sqrt{3}a/2]^{\text{T}}$, and $\bm{\delta}_{\text{CA}} = [-a/2,-\sqrt{3}a/2]^{\text{T}}$.
The eigenvalues of this model are given by
\begin{equation}
    0, \quad \frac{3}{2}\left( \alpha^{2} + \frac{1}{\alpha^{2}} \right) \pm \sqrt{ \frac{1}{4}\left(\alpha^{4} + \frac{1}{\alpha^{4}} \right) + \frac{5}{2} + 2\left[ \cos(k_{x}a+\sqrt{3}k_{y}a) + \cos(k_{x}a-\sqrt{3}k_{y}a) + \cos(2k_{x}a) \right] }.
\end{equation}
The (unnormalized) eigenfunction of the isolated flat band is simply given as
\begin{equation}
    \mqty[
    (1/\alpha)^{2}e^{-\mathrm{i}\bm{k}\cdot\bm{\delta}_{\text{BC}}} - \alpha^{2}e^{\mathrm{i}\bm{k}\cdot\bm{\delta}_{\text{BC}}} \\
    -(1/\alpha)^{2}e^{\mathrm{i}\bm{k}\cdot\bm{\delta}_{\text{CA}}} + \alpha^{2}e^{-\mathrm{i}\bm{k}\cdot\bm{\delta}_{\text{CA}}} \\
    -e^{\mathrm{i}\bm{k}\cdot\bm{\delta}_{\text{AB}}} + e^{-\mathrm{i}\bm{k}\cdot\bm{\delta}_{\text{AB}}}
    ].
\end{equation}

\noindent
\textit{Kagome-$3$ model}.---The third example is the Kagome-$3$ model~\cite{Bergman2008} in Fig.~1(c), which has two-fold degenerate isolated flat bands and one dispersive band.
The Hamiltonian $H_{\text{K}3\bm{k}}$ is given by the following.
\begin{equation}
    H_{\text{K}3\bm{k}} = 2t\cdot\mqty[
    \cos(\bm{k}\cdot\bm{c}_{\text{AA}})+1 & \cos(\bm{k}\cdot\bm{a}_{\text{AB}}) + \cos(\bm{k}\cdot\bm{b}_{\text{AB}}) & \cos(\bm{k}\cdot\bm{a}_{\text{CA}}) + \cos(\bm{k}\cdot\bm{b}_{\text{CA}}) \\
    \cos(\bm{k}\cdot\bm{a}_{\text{AB}}) + \cos(\bm{k}\cdot\bm{b}_{\text{AB}}) & \cos(\bm{k}\cdot\bm{c}_{\text{BB}})+1 & \cos(\bm{k}\cdot\bm{a}_{\text{BC}}) + \cos(\bm{k}\cdot\bm{b}_{\text{BC}}) \\
    \cos(\bm{k}\cdot\bm{a}_{\text{CA}}) + \cos(\bm{k}\cdot\bm{b}_{\text{CA}}) & \cos(\bm{k}\cdot\bm{a}_{\text{BC}}) + \cos(\bm{k}\cdot\bm{b}_{\text{BC}}) & \cos(\bm{k}\cdot\bm{c}_{\text{CC}})+1
    ],
\end{equation}
where $\bm{a}_{\text{AB}}=\mqty[a,0]^{\text{T}}$, $\bm{a}_{\text{BC}}=\mqty[-a/2, \sqrt{3}a/2]^{\text{T}}$, 
$\bm{a}_{\text{CA}}=\mqty[-a/2,-\sqrt{3}a/2]^{\text{T}}$,
$\bm{b}_{\text{AB}}=\mqty[0,-\sqrt{3}a]^{\text{T}}$,
$\bm{b}_{\text{BC}}=\mqty[3a/2, \sqrt{3}a/2]^{\text{T}}$,
$\bm{b}_{\text{CA}}=\mqty[-3a/2,\sqrt{3}a/2]^{\text{T}}$,
$\bm{c}_{\text{AB}}=\mqty[-a,\sqrt{3}a]^{\text{T}}$,
$\bm{c}_{\text{BC}}=\mqty[-a,-\sqrt{3}a]^{\text{T}}$,
$\bm{c}_{\text{CA}}=\mqty[2a,0]^{\text{T}}$.
The eigenvalues of the model are
\begin{equation}
    0 \ (\text{two-fold}),\quad 6+\cos(2k_{x}a)+\cos(k_{x}a+\sqrt{3}k_{y}a)+\cos(k_{x}a-\sqrt{3}k_{y}a).
\end{equation}
The two (unnormalized) eigenfunctions of the two-fold degenerate isolated flat bands are given by the following expressions.
\begin{equation}
    \mqty[
    -(1+\cos(2k_{x}a)) \\
    0 \\
    \cos(3k_{x}a/2 - \sqrt{3}k_{y}a/2) + \cos(k_{x}a/2 + \sqrt{3}k_{y}a/2)
    ], \quad \mqty[
    -( \cos(k_{x}a) + \cos(\sqrt{3}k_{y}a) ) \\
    2 + \cos(2k_{x}a) + \cos(k_{x}a-\sqrt{3}k_{y}a) \\
    -2\cos(k_{x}a)\cos(k_{x}a/2 + \sqrt{3}k_{y}a/2)
    ].
\end{equation}

\noindent
To check the validity of our formula, we use the analytical expressions of the eigenfunctions of the isolated flat bands to avoid the numerical derivatives.
We use the numerical values of the eigenfunctions for the dispersive bands.
We numerically calculate $\Lambda$ for each lattice model with the aid of the analytic form of the eigenfunctions of the isolated flat bands to obtain the solid lines in Fig.~(2) in the main text.
We normalize the responses to plot the markers in Fig.~2 obtained from Eqs.~(\ref{eq:TR}) and~(\ref{eq:TH}) by calculating the maximum value of $|\sigma^{\text{Total}}_{xy}/\sigma_{xy0}|$.

\section*{Plots of the geometric part of the formula}
We present the plots of the geometric part of the formula: $\sum_{l,m\in F}\text{Re}[ (Q^{lm}_{xy\bm{k},n} - Q^{lm}_{yx\bm{k},n})Q^{ml}_{xy\bm{k}}]$ of the models we used.
The quantity $\Lambda$ in the main text is derived by multiplying the band gap $(E_{n\bm{k}}-E_{\text{flat}})$ for each point in the first Brillouin zone, integrating about each $\bm{k}$, and multiplying the factor $-4$.
For the honeycomb lattice model, $D=\{1\}$ and $F=\{2\}$ hold, and the geometric part returns to the square of the Berry curvature of the isolated flat band, as mentioned in the main text.
For the Gapped Kagome lattice model, $D=\{1,2\}$ and $F=\{3\}$ hold, and there are two different contributions from each dispersive band.
The contribution from the upper band is smaller than that from the lower band, despite the increase in band gap.
For the Kagome-$3$ model, $D=\{1\}$ and $F=\{2,3\}$ hold, and there is only one contribution from the dispersive band.
We note that $Q^{11}_{xy\bm{k},1}=Q^{11}_{yx\bm{k},1}$ and $Q^{22}_{xy\bm{k},1}=Q^{22}_{yx\bm{k},1}$ hold, and the contributions from the isolated flat bands with the same labels are zero.
Hence, the geometric contributions only come from the transition between the isolated flat bands with the different labels: $Q^{12}_{xy\bm{k},1}$, $Q^{12}_{yx\bm{k},1}$, $Q^{21}_{xy\bm{k},1}$, and $Q^{21}_{yx\bm{k},1}$.

\begin{figure}
    \centering
    \includegraphics[scale=1.0]{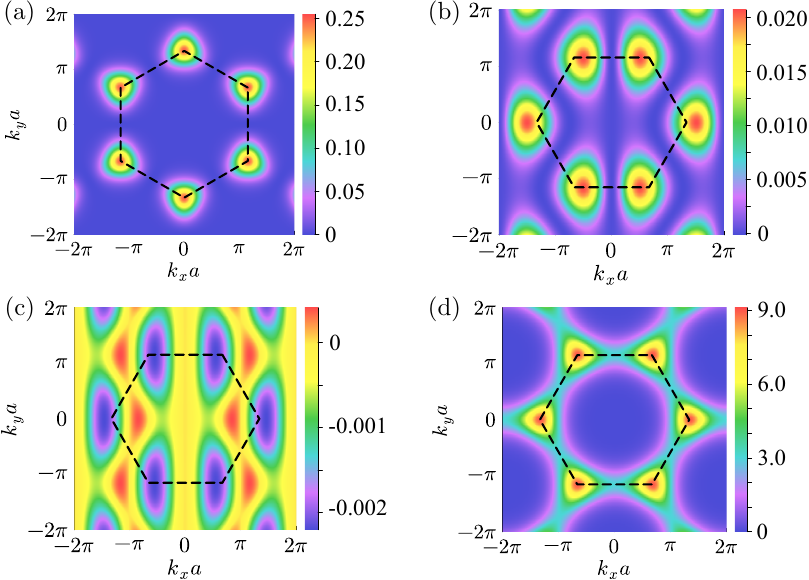}
    \caption{Plots of the geometric part of the formula of the models in the main text.
    The dashed lines represent the boundary of the first Brillouin zone.
    (a) Honeycomb lattice model.
    (b) Gapped Kagome lattice model for the lower dispersive band ($n=1$).
    (c) Gapped Kagome lattice model for the upper dispersive band ($n=2$).
    (d) Kagome-$3$ lattice model.}
    \label{Fig_supple_geometric}
\end{figure}

\section*{Transport and thermodynamic contributions to the Hall conductivity in dispersive band systems}
\begin{figure}
    \centering
    \includegraphics[scale=0.6]{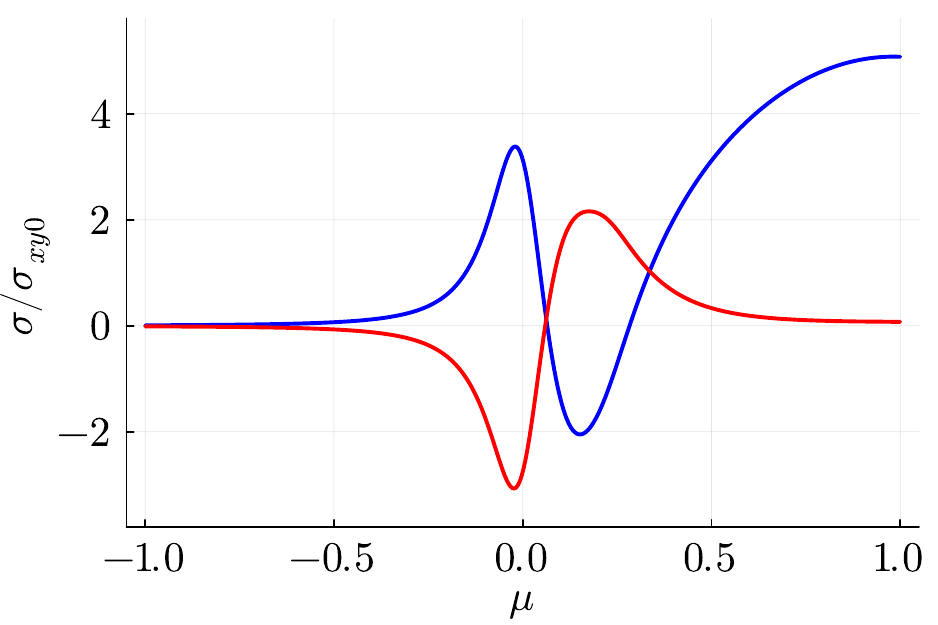}
    \caption{Transport ($\sigma_{xy}^{\text{TR}}$, blue) and thermodynamic ($\sigma_{xy}^{\text{TH}}$, red) contributions to the Hall conductivity for the honeycomb lattice model with $m=3.0$, $T=0$, $\delta t=0.2$, and $\Gamma=0.1$.
    }
    \label{Fig_supple}
\end{figure}

We show how the transport and thermodynamic contributions to the Hall conductivity behave as the isolated flat band turns into a dispersive one.
When the isolated flat band becomes dispersive, the transport contribution becomes dominant, and the thermodynamic contribution is less important.
In Fig.~\ref{Fig_supple}, the blue and red curves correspond to the transport ($\sigma_{xy}^{\text{TR}}$) and thermodynamic ($\sigma_{xy}^{\text{TH}}$) contributions of the Hall conductivity.
The parameters are set to $m=3.0$, $T=0$, $\delta t=0.2$, and $\Gamma=0.1$.
The isolated flat band is originally located at $\mu=0$.

\section*{Modification of the formula by small finite dispersion of the flat band ($N=M=1$)}
Let us consider the two-band system with one dispersive band (band $1$, $E_{1\bm{k}}$) and one isolated flat band (band $2$, $E_{2\bm{k}}=\text{const}.$), satisfying $\text{min}(|E_{1\bm{k}}-E_{2\bm{k}}|)\gg \Gamma>0$ for any $\bm{k}$ in the Brillouin zone.
We put $E_{2\bm{k}}=0$ for simplicity.
We then introduce a finite small dispersion $\delta E_{2\bm{k}}$ to the band $2$, satisfying $\text{max}(|\delta E_{2\bm{k}}|)\ll\text{min}(|E_{1\bm{k}} - E_{2\bm{k}}|)$.
We only pick up the first order terms of $\delta E_{2\bm{k}}$, neglect higher order terms, and consider the finite temperature case.
The modified Hall conductivity $\sigma_{xy}(\mu)$ as a function of chemical potential $\mu$ (measured from $E_{2\bm{k}}=0$) is given by the following expression.
\begin{align}
    &\frac{\sigma_{xy}^{\text{Total}}(\mu)}{\sigma_{xy0}} \notag \\
    =&\pi\frac{1}{N}\sum_{\bm{k}}\Omega^{2}_{xy,\bm{k}}(E_{1\bm{k}}-E_{2\bm{k}})\frac{\beta}{4}\sech^{2}\left[ \frac{\beta}{2}(\delta E_{2\bm{k}} - \mu) \right] \notag \\
    & -3\pi\frac{1}{N}\sum_{\bm{k}}\sum_{\nu,\lambda}g_{\bm{k}}g^{\nu\lambda}_{\bm{k}}f_{\nu\lambda,\bm{k}}\frac{\beta}{4}\sech^{2}\left[ \frac{\beta}{2}(\delta E_{2\bm{k}} - \mu) \right] \notag \\
    & +\frac{3\pi}{2}\frac{1}{N}\sum_{\bm{k}}\left[ (\partial_{x}\delta E_{2\bm{k}})(\partial_{x} g_{yy}) + (\partial_{y}\delta E_{2\bm{k}})(\partial_{y}g_{xx}) + 2(\partial_{x}\partial_{y}\delta E_{2\bm{k}})g_{xy} \right]\frac{\beta}{4}\sech^{2}\left[ \frac{\beta}{2}(\delta E_{2\bm{k}} - \mu) \right],
\end{align}
where $\Omega_{xy,\bm{k}}$ is the Berry curvature of the band $2$ without the finite dispersion defined as $\Omega_{xy,\bm{k}}:=2\text{Im}\braket{\partial_{x}u_{2\bm{k}}}{\partial_{y}u_{2\bm{k}}}$, $g_{\nu\lambda,\bm{k}}$ is the quantum metric of the band $2$ defined as
\begin{equation}
    g_{\nu\lambda,\bm{k}} = \frac{1}{2}\big[ \braket{\partial_{\nu}u_{2\bm{k}}}{\partial_{\lambda}u_{2\bm{k}}} + \braket{\partial_{\lambda}u_{2\bm{k}}}{\partial_{\nu}u_{2\bm{k}}} \big] - \braket{\partial_{\nu}u_{2\bm{k}}}{u_{2\bm{k}}}\braket{u_{2\bm{k}}}{\partial_{\lambda}u_{2\bm{k}}},
\end{equation}
which is real.
The second term is written in a (little bit) concise form by introducing
\begin{align}
    g_{\bm{k}} &= \text{det}g_{\nu\lambda,\bm{k}}, \\
    g^{\nu\lambda}_{\bm{k}} &= (g_{\nu\lambda,\bm{k}})^{-1} = \frac{1}{g_{\bm{k}}}\mqty[ g_{yy,\bm{k}} & -g_{xy,\bm{k}} \\ -g_{xy,\bm{k}} & g_{xx,\bm{k}}], \\
    f_{\nu\lambda, \bm{k}} &= \frac{1}{E_{1\bm{k}} - E_{2\bm{k}}}\mqty[ (\partial_{x}E_{1\bm{k}})(\partial_{x}\delta E_{2\bm{k}}) & (\partial_{x}E_{1\bm{k}})(\partial_{y}\delta E_{2\bm{k}}) \\ (\partial_{y}E_{1\bm{k}})(\partial_{x}\delta E_{2\bm{k}}) & (\partial_{y}E_{1\bm{k}})(\partial_{y}\delta E_{2\bm{k}})],
\end{align}
meaning
\begin{align}
    &g_{\bm{k}}g^{\nu\lambda}_{\bm{k}}f_{\nu\lambda,\bm{k}} \notag \\
    =&\frac{ (\partial_{x}E_{1\bm{k}})(\partial_{x}\delta E_{2\bm{k}})g_{yy,\bm{k}} + (\partial_{y}E_{1\bm{k}})(\partial_{y}\delta E_{2\bm{k}})g_{xx,\bm{k}} - (\partial_{x}E_{1\bm{k}})(\partial_{y}\delta E_{2\bm{k}})g_{xy,\bm{k}} - (\partial_{y}E_{1\bm{k}})(\partial_{x}\delta E_{2\bm{k}})g_{xy,\bm{k}}}{E_{1\bm{k}} - E_{2\bm{k}}}.
\end{align}
The mofified formula has a temperature dependence only by $(\beta/4)\sech^{2}(\beta(\delta E_{2\bm{k}} - \mu)/2)$.
If $\text{max}(|\beta\cdot \delta E_{2\bm{k}}|)\ll1$ holds, the temperature-dependent part does not depend on the wavenumber $\bm{k}$ effectively, and the total Hall conductivity becomes symmetric around $\mu=E_{2\bm{k}} =0$, meaning that the finite dispersion only modifies the magnitude of the Hall conductivity.
The condition $\text{max}(|\beta\cdot \delta E_{2\bm{k}}|)\ll1$ can be interpereted that the finite dispersion $\delta E_{2\bm{k}}$ is ``buried" by the thermal energy determined by $k_{\text{B}}T$ and the band $2$ is effectively seen to be flat even though it has a finite dispersion.
If the temperature is sufficiently small and the finite dispersion is not negligible (the condition $\text{max}(|\beta\cdot \delta E_{2\bm{k}}|)\ll1$ is not satisfied anymore), the peak position of the Hall conductivity is shifted from $\mu=E_{2\bm{k}}=0$ and the structure is not simple at all.
A similar argument for $T=0$ and finite $\Gamma>0$ holds by interpreting $k_{\text{B}}T\to\Gamma$.\\

The procedure for deriving the modified formula is no different from the method for deriving the original formula.
What we need to do is modify the explicit expressions of $H^{\nu}_{\bm{k}}$, $H^{xy}_{\bm{k}}$:
\begin{align}
    H^{\nu}_{\bm{k}} &= \mqty[ H^{\nu}_{11\bm{k}} & H^{\nu}_{12\bm{k}} \\ H^{\nu}_{21\bm{k}} & H^{\nu}_{22\bm{k}}], \notag \\
    H^{\nu}_{11\bm{k}} &= \partial_{\nu}E_{1\bm{k}}, \notag \\
    H^{\nu}_{22\bm{k}} &= \partial_{\nu}\delta E_{2\bm{k}}, \notag \\
    H^{\nu}_{12\bm{k}} &= [H^{\nu}_{21\bm{k}}]^{*} = -(E_{1\bm{k}} - \delta E_{2\bm{k}})\braket{u_{1\bm{k}}}{\partial_{\nu}u_{2\bm{k}}}, \notag
\end{align}
\begin{align}
    H^{xy}_{\bm{k}} &= \mqty[ H^{xy}_{11\bm{k}} & H^{xy}_{12\bm{k}} \\ H^{xy}_{21\bm{k}} & H^{xy}_{22\bm{k}}], \notag \\
    H^{xy}_{11\bm{k}} &= (\partial_{x}\partial_{y}E_{1\bm{k}}) - 2(E_{1\bm{k}} - \delta E_{2\bm{kl}})g_{xy,\bm{k}}, \notag \\
    H^{xy}_{22\bm{k}} &= (\partial_{x}\partial_{y}\delta E_{2\bm{k}}) + 2(E_{1\bm{k}} - \delta E_{2\bm{k}})g_{xy,\bm{k}}, \notag \\
    H^{xy}_{12\bm{k}} &= H^{xy}_{21\bm{k}} \notag \\
    &= -(\partial_{x}E_{1\bm{k}})\braket{u_{1\bm{k}}}{\partial_{y}u_{2\bm{k}}} - (\partial_{y}E_{1\bm{k}})\braket{u_{1\bm{k}}}{\partial_{x}u_{2\bm{k}}} \notag \\
    &\quad +(\partial_{x}\delta E_{2\bm{k}})\braket{u_{1\bm{k}}}{\partial_{y}u_{2\bm{k}}} + (\partial_{y}\delta E_{2\bm{k}})\braket{u_{1\bm{k}}}{\partial_{x}u_{2\bm{k}}} \notag \\
    &\quad - (E_{1\bm{k}} - \delta E_{2\bm{k}})\braket{\partial_{y}u_{1\bm{k}}}{u_{2\bm{k}}}\braket{u_{2\bm{k}}}{\partial_{x}u_{2\bm{k}}} \notag \\
    &\quad - (E_{1\bm{k}} - \delta E_{2\bm{k}})\braket{\partial_{x}u_{1\bm{k}}}{u_{2\bm{k}}}\braket{u_{2\bm{k}}}{\partial_{y}u_{2\bm{k}}} \notag \\
    &\quad - (E_{1\bm{k}} - \delta E_{2\bm{k}})\braket{u_{1\bm{k}}}{\partial_{x}\partial_{y}u_{2\bm{k}}},
\end{align}
and the Green's functions:
\begin{align}
    G^{\text{R}}_{\bm{k}} &= \mqty[ (\varepsilon + \mathrm{i}\Gamma - E_{1\bm{k}})^{-1} & 0 \\ 0 & (\varepsilon + \mathrm{i}\Gamma - \delta E_{2\bm{k}})^{-1}] = \mqty[ G^{\text{R}}_{1\bm{k}} & 0 \\ 0 & G^{\text{R}}_{2\bm{k}}], \notag \\
    G^{\text{A}}_{\bm{k}} &= \mqty[ (\varepsilon - \mathrm{i}\Gamma - E_{1\bm{k}})^{-1} & 0 \\ 0 & (\varepsilon - \mathrm{i}\Gamma - \delta E_{2\bm{k}})^{-1}] = \mqty[ G^{\text{A}}_{1\bm{k}} & 0 \\ 0 & G^{\text{A}}_{2\bm{k}}]. \notag
\end{align}
The approximation we use is the following.
\begin{align}
    G^{\text{R}}_{1\bm{k}} &\approx -\frac{(E_{1\bm{k}} - E_{2\bm{k}})+\mathrm{i}\Gamma}{(E_{1\bm{k}} - E_{2\bm{k}})^{2}}, \notag \\
    G^{\text{A}}_{1\bm{k}} &\approx -\frac{(E_{1\bm{k}} - E_{2\bm{k}})-\mathrm{i}\Gamma}{(E_{1\bm{k}} - E_{2\bm{k}})^{2}}, \notag \\
    G^{\text{R}}_{2\bm{k}} &\approx \frac{(\varepsilon - \delta E_{2\bm{k}})-\mathrm{i}\Gamma}{(\varepsilon - \delta E_{2\bm{k}})^{2}+\Gamma^{2}}, \notag \\
    G^{\text{A}}_{2\bm{k}} &\approx \frac{(\varepsilon - \delta E_{2\bm{k}})+\mathrm{i}\Gamma}{(\varepsilon - \delta E_{2\bm{k}})^{2}+\Gamma^{2}}. \notag
\end{align}
Various terms appear during the calculation, but many of them do not contribute to the final results by taking the imaginary parts and the limit of $\Gamma\to 0$.
The relevant terms include one of the following.
\begin{align}
    \frac{\Gamma}{(\varepsilon - \delta E_{2\bm{k}})^{2}+\Gamma^{2}}&\xrightarrow{\Gamma\to 0}\pi \delta(\varepsilon - \delta E_{2\bm{k}}), \notag \\
    \frac{\Gamma^{3}}{[ (\varepsilon - \delta E_{2\bm{k}})^{2}+\Gamma^{2} ]^{2}}&\xrightarrow{\Gamma\to 0}\frac{\pi}{2} \delta(\varepsilon - \delta E_{2\bm{k}}). \notag
\end{align}
By summarizing the remaining terms, we obtain the modified formula.

\end{document}